\documentclass[twocolumn,showpacs,amsmath,amssymb,pr,superscriptaddress]{revtex4-2}
\usepackage{graphicx}
\usepackage{epsf}
\usepackage{ulem}
\usepackage{color}
\usepackage[unicode=true,colorlinks=true,urlcolor=blue,citecolor=blue]{hyperref}
\usepackage{xcolor}

\begin{document}

\title{Magnetophonon Resistance Oscillations in Structures with a GaAs Quantum Well and Barriers of AlAs/GaAs$\langle\delta$-Si$\rangle$ Superlattices}

\author{I. L. Drichko}
\affiliation{Ioffe Physical-Technical Institute, St Petersburg, Russia}
\author{I. Yu. Smirnov}
\affiliation{Ioffe Physical-Technical Institute, St Petersburg, Russia}
\author{M. O. Safonchik}
\affiliation{Ioffe Physical-Technical Institute, St Petersburg, Russia}
\author{M. A. Shakhov}
\affiliation{Ioffe Physical-Technical Institute, St Petersburg, Russia}
\author{A. K. Bakarov}
\affiliation{Rzhanov Institute of Semiconductor Physics, Novosibirsk, Russia}
\author{A. A. Bykov}
\affiliation{Rzhanov Institute of Semiconductor Physics, Novosibirsk, Russia}

\begin{abstract}
Magnetophonon resistance oscillations (MPR) associated with the resonant scattering of electrons by optical phonons at temperatures of 77--240 K, as well as resonant scattering of electrons by acoustic phonons (PIRO) at temperatures of 10--25 K, were investigated in the same samples featuring a GaAs quantum well and AlAs/GaAs superlattice barriers doped with Si. The study of MPR demonstrated that resonant electron scattering occurs on bulk longitudinal optical phonons and does not depend on the dimensionality of the system or inter-subband transitions in systems with two subbands of size quantization. However, the amplitude of the oscillation with number $N=1$ in two-dimensional structures depends on the interplay of scattering mechanisms, which, in turn, is influenced by the structure of the system. As for PIRO, in samples with two size quantization subbands, resonant electron scattering by longitudinal acoustic phonons is observed against the background of inter-subband transitions (MISO), leading to their interference.
\end{abstract}

\maketitle

\section{MAGNETOPHONON RESONANCE OF RESISTANCE ON OPTICAL PHONONS}

\subsection{Introduction}

\begin{figure}[h]
\centering
\includegraphics[width=0.9\columnwidth]{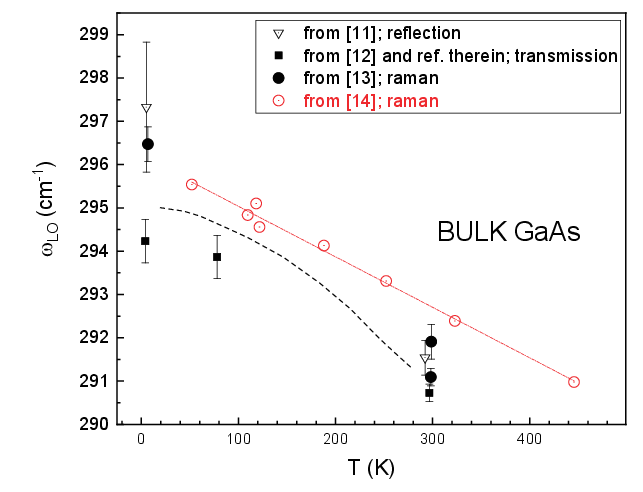} 
\caption{Temperature dependences of longitudinal optical phonon frequencies, measured by different authors and different methods (IR reflection spectroscopy~\cite{Hass1962}, IR transmission spectroscopy~\cite{Stradling1968} and references therein, Raman spectroscopy~\cite{Mooradian1966,Jusserand1981}).}
\label{fig1}
\end{figure}

In 1961, Gurevich and Firsov~\cite{Gurevich1961} theoretically predicted magnetophonon resonance in three-dimensional semiconductors arising from resonant electron scattering by optical phonons. As early as 1963, the effect was experimentally observed in $n$-InSb crystals by the authors of Refs.~\cite{Puri1963,Shalyt1964}. Nearly two decades later, in 1980, magnetophonon resonance (MPR) due to optical phonons was also discovered in two-dimensional electron systems, including single heterojunctions and GaAs/AlGaAs superlattices~\cite{Tsui1980}.
Following its discovery, MPR was extensively investigated in a variety of systems, particularly in GaAs quantum wells~\cite{Tsui1982,Kido1982,Brummell1987,Leadley1994,Faugeras2004,Nicholas1991}. In all of these studies, the magnetic-field positions of the magnetoresistance oscillation maxima, $\rho_{xx}(B)$, were determined experimentally. Since the oscillation maxima are expected to satisfy the resonance condition
\begin{equation}
\omega_{LO} = N \omega_c, \quad \omega_c = \frac{eB}{m^*c},
\end{equation}
where $\omega_{LO}$ is the optical phonon frequency, $\omega_c$ is the cyclotron frequency, $m^*$ is the effective mass, and $N$ is the oscillation number, then some authors~\cite{Tsui1982} used experimentally determined values of these magnetic fields to calculate the effective mass, substituting the optical phonon frequency $\omega_{LO}=293\ \text{cm}^{-1}$. If we consider how the values of the optical phonon frequency determined by different authors and different methods are related (Fig.~\ref{fig1}), it becomes clear that the effective mass value determined in this way depends on the choice of the optical phonon frequency.

The authors of Refs.~\cite{Tsui1980,Tsui1982,Kido1982,Brummel1987}, on the contrary, calculated the frequency of optical phonons using the value of the effective mass determined from cyclotron resonance. It should be noted that in all these works the values of the magnetic field $B$ for the position of the oscillation peak with $N=1$ were the same and, within the error limits, were equal to $B=(23\pm 1.5)$ T.

The theory of MPR on optical phonons for a heterojunction was constructed in Ref.~\cite{Lassnigt1983}. The calculation showed that the amplitude of resistance oscillations $\Delta\rho_{\rm xx}$ associated with MPR in a GaAs-AlGaAs heterojunction should be approximately 20\% of $\rho_{\rm xx}$. However, in all experiments the amplitude of MPR oscillations was small, $\Delta\rho_{\rm xx}/\rho_{\rm xx}<1\%$, and was determined using complex computer processing.

For two-dimensional objects of the GaAs/AlGaAs type with one size-quantization subband, the theory of magnetophonon resonance was constructed in Ref.~\cite{Afonin2000}, in which the electron-electron interaction was taken into account. The result of the work was the assertion that the resonant scattering of electrons in a two-dimensional structure occurs at the optical phonon frequency, which differs from the frequency of the bulk longitudinal optical phonon and is shifted to the frequency of the transverse one. However, it follows from the experimental work~\cite{Stradling1968} that the maximum of resistance oscillations with $N=1$ and in a three-dimensional sample also occurs in a magnetic field, the strength of which fits into the same limits as for two-dimensional systems. Since it was shown that resonant scattering of electrons in crystals occurs on bulk longitudinal optical phonons, it can be assumed that in two-dimensional objects resonant scattering is also carried out on these phonons.

If in works~\cite{Tsui1982,Kido1982,Brummel1987,Leadley1994,Faugeras2004,Nicholas1991} the studies of magnetophonon resonance were carried out on GaAs quantum wells with one filled energy subband of size quantization, then in the present work we report on the observation of magnetophonon resonances on optical phonons in single GaAs quantum wells with two filled energy subbands.

\subsection{Experimental results and their discussion}

\begin{figure}[t]
\centering
\includegraphics[width=\columnwidth]{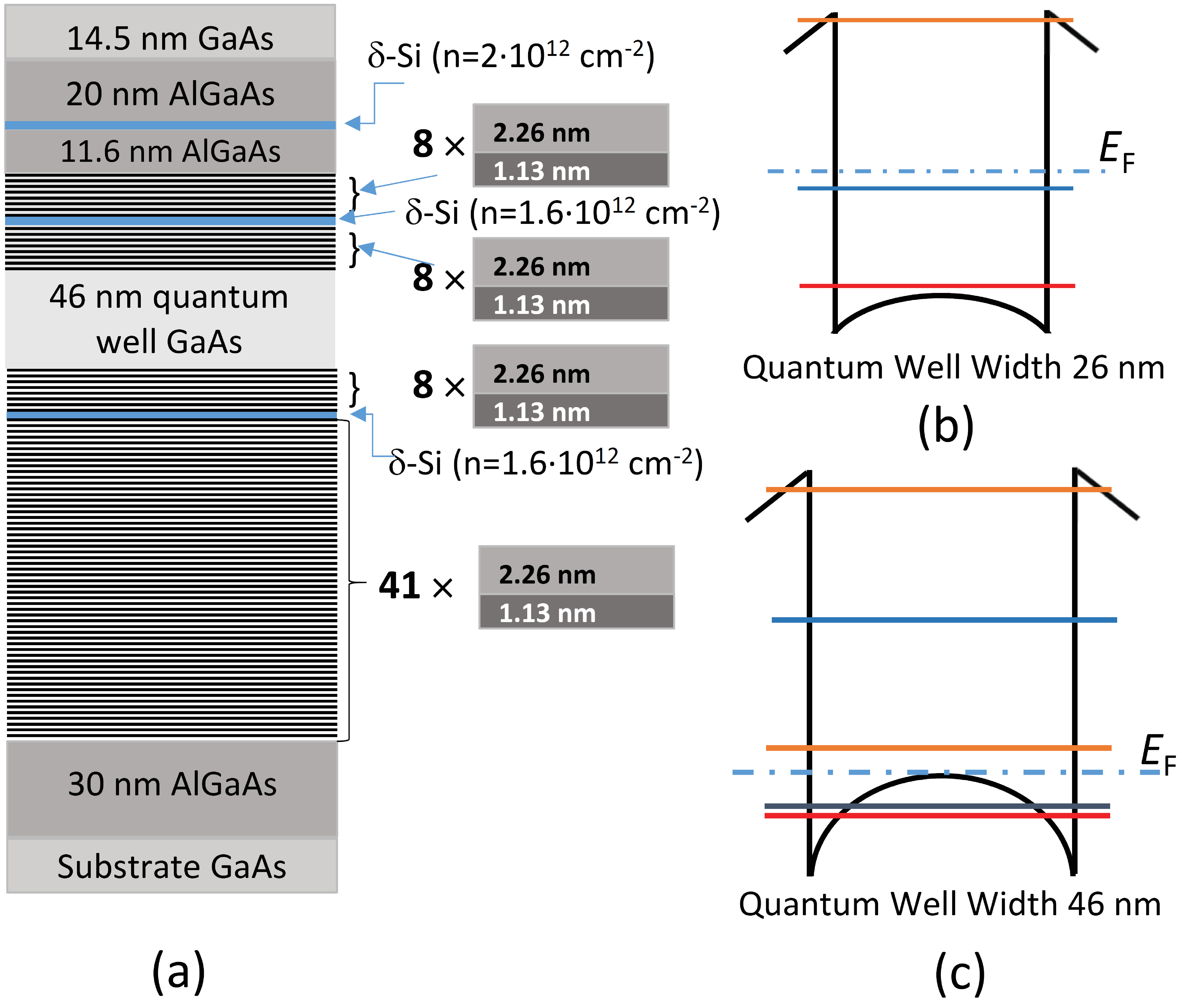} 
\caption{(a) Construction of the structures (the same for both samples). The samples differed only in the well width: in sample 1 the well width was 26 nm, and in the second 46 nm. Energy schemes for samples 1 (b) and 2 (c).}
\label{fig2}
\end{figure}

The objects of the study were two structures with quantum wells (QW) $n$-GaAs and barriers made of short-period AlAs/GaAs superlattices~\cite{Bykov2020}.

Despite the same electron concentration and mobility in quantum wells ($n=8.2\cdot 10^{11}\ \text{cm}^{-2}$, $\mu\sim 10^6\ \text{cm}^2/\text{V}\cdot\text{s}$) in the studied samples, the different widths of the wells determined the different structure of their energy bands: in the first sample, two size quantization subbands with concentrations $6.2\cdot 10^{11}\ \text{cm}^{-2}$ and $1.9\cdot 10^{11}\ \text{cm}^{-2}$, separated by an energy gap, arose $\Delta_{SAS}=15.5$ meV~\cite{Goran2009}. In the second sample with a wide (46 nm) quantum well, a two-layer system was formed due to the Coulomb repulsion of electrons to the heterointerfaces. Such a system is also characterized by a two-subband energy spectrum, although the energy of subband separation in this case is much less than in the first sample, $\Delta_{SAS}=1.45$ meV, and the electron concentrations in the subbands differed little: $4.27\cdot 10^{11}\ \text{cm}^{-2}$ and $3.93\cdot 10^{11}\ \text{cm}^{-2}$.

\begin{figure}[h]
\centering
\includegraphics[width=0.9\columnwidth]{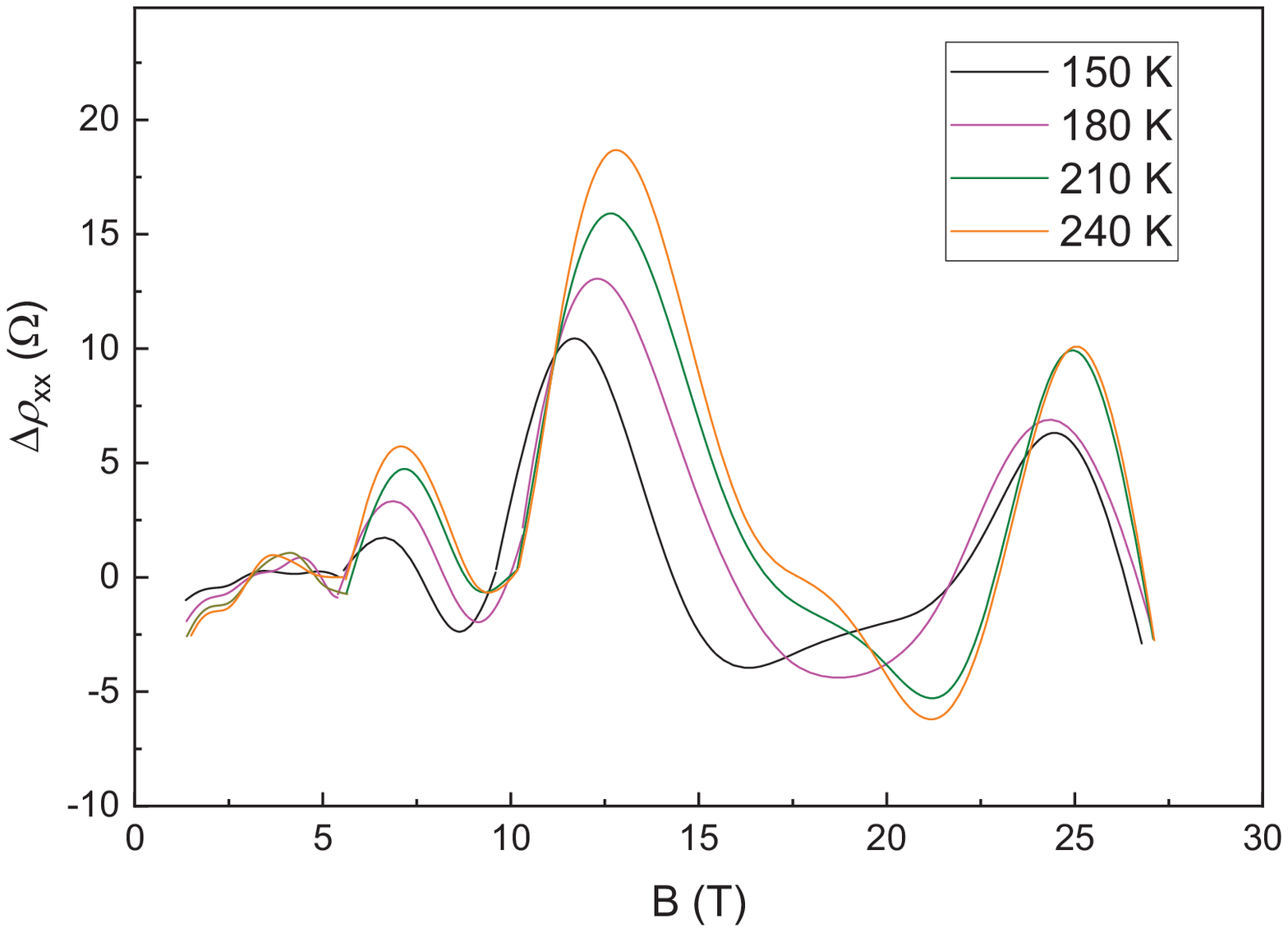} 
\caption{Magnetophonon oscillations in sample 1 at different temperatures.}
\label{fig3}
\end{figure}

\begin{figure}[h]
\centering
\includegraphics[width=\columnwidth]{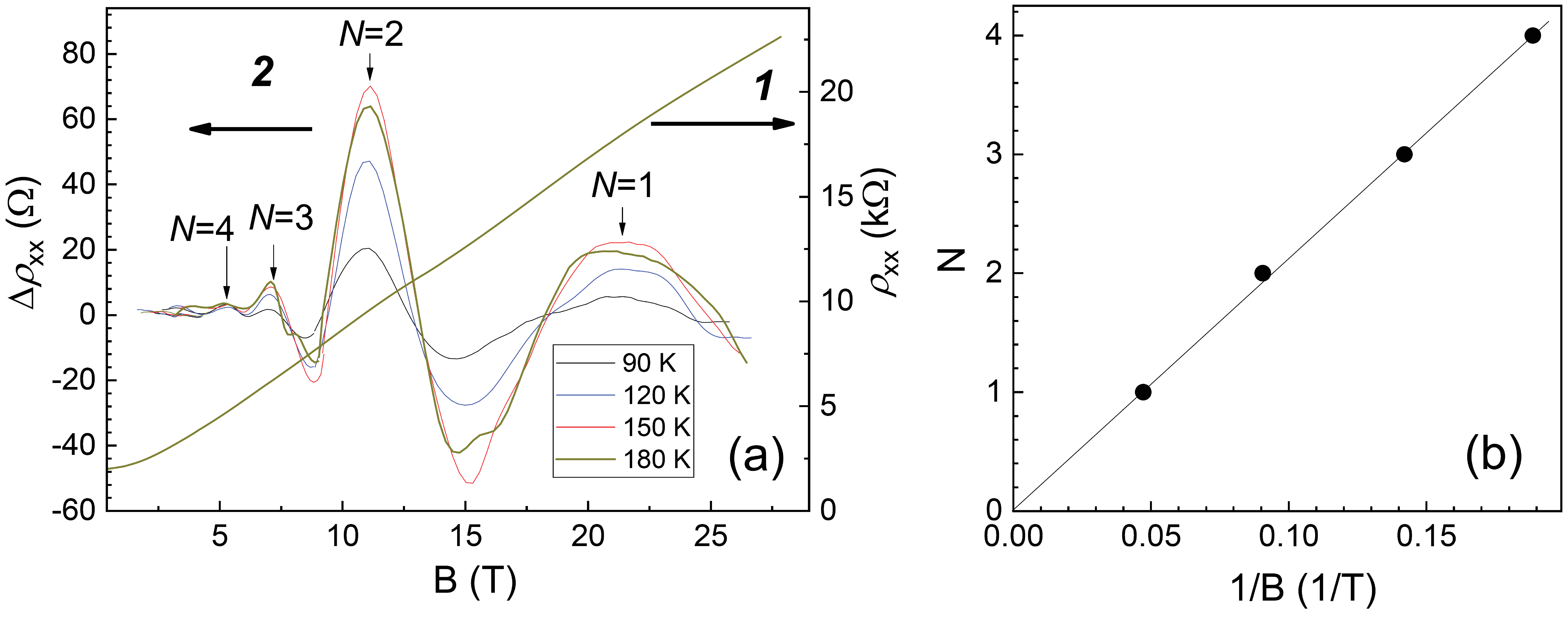} 
\caption{(a) Magnetophonon oscillations in sample 2 at different temperatures: 1 -- experimental dependence $\rho_{xx}(B)$ at $T = 180$ K; 2 -- dependences of the oscillating part $\Delta \rho_{xx}(B)$ at different temperatures; arrows show the numbers of the magnetophonon resonance peaks. (b) Dependence of the oscillation number on the inverse magnetic field, demonstrating their periodicity $1/B$.}
\label{fig4}
\end{figure}

The measurements were carried out in a pulsed magnetic field of up to 30 T, perpendicular to the plane of the sample, with a duration of 12 ms at $T\approx 77-240$ K at a constant (for the duration of the magnetic field) current, the value of which did not exceed 1 mA. In both samples, resistance oscillations $\rho_{\rm xx}$ were observed depending on the magnetic field (Figs.~\ref{fig3} and \ref{fig4}a). The maximum oscillation amplitude was observed at $T\sim 240$ K for sample 1 and at $T\sim 150$ K for sample 2.

The dependence of magnetoresistance on the magnetic field in the oscillation region can be described by the formula~\cite{Stradling1968}
\begin{equation}
\frac{\Delta \rho_{xx}}{\rho_{xx}} \propto \cos \left( \frac{2 \pi \omega_{LO}}{\omega_{c}} \right) \exp \left( -\frac{\gamma \omega_{LO}}{\omega_{c}} \right).
\end{equation}
This formula was obtained for a three-dimensional sample, but it turned out to be applicable for the two-dimensional case as well~\cite{Kido1982,Brummel1987,Leadley1994}. The value $\gamma$ does not depend on the magnetic field, but depends on the temperature and mobility of carriers in the sample. Indeed, the value $\gamma$ turned out to be different for different samples. In our samples, $\gamma \sim 1.5$.

The position of the peak of the oscillations with $N = 1$ in sample 2 corresponds to a magnetic field $B = 22$ T, i.e. the same region of magnetic fields that were observed in Refs.~\cite{Tsui1982,Kido1982,Brummel1987,Leadley1994,Faugeras2004,Nicholas1991}. This fact confirms that it is precisely magnetophonon oscillations that are observed, caused by the resonant interaction of electrons with longitudinal optical phonons.

The oscillation maxima in sample 1 in Fig.~\ref{fig3} are slightly shifted in magnetic fields to the larger side, which we explain by the smallness of the effect and, possibly, the influence of the second subband of size quantization.

According to formula (2), the oscillation amplitude should increase with increasing magnetic field. However, if the amplitudes of the last three ($N = 2, 3, 4$) oscillations grow with increasing magnetic field as $\exp(-\gamma \omega_{LO}/\omega_{c})$, then the oscillation amplitude with $N = 1$ ($B \approx 22$ T) did not only increase compared to the oscillation amplitude with $N = 2$, but was even smaller than it in both samples. This problem was encountered by everyone who studied magnetophonon resonance in two-dimensional GaAs structures. In order to understand the cause of this phenomenon, the effect of scattering mechanisms on the amplitude of magnetophonon oscillations in high and low magnetic fields was experimentally and theoretically investigated in Ref.~\cite{Leadley1994}. It was shown that in high magnetic fields with weak elastic scattering, resonant scattering by optical phonons sharply increases the width of the Landau level, which leads to a decrease in the density of states and, accordingly, to a decrease in the oscillation amplitude with $N = 1$ and even to its disappearance. In low magnetic fields, this effect is absent. The experimental part of the work~\cite{Leadley1994} demonstrates the dependence of the amplitude of the first maximum of MPR on the spacer width. The larger the spacer width, the smaller the amplitude of the first maximum. Indeed, with an increase in the spacer width, the influence of elastic scattering on the width of Landau levels decreases and the role of inelastic scattering on optical phonons becomes predominant.

Now let us consider how the structures under study are arranged. It was mentioned above that in the structures under study the barriers to the $n$-GaAs quantum wells were made of short-period GaAs/AlAs superlattices. The source of free electrons in such a heterostructure were two $\delta$-doped GaAs layers, which were located in the superlattice barriers on both sides of the $n$-GaAs quantum well at a distance of 29.4 nm from its boundaries. In this case, the suppression of the elastic scattering of electrons on the random potential of ionized donors is achieved not only by spatial separation of the doping and transport regions, but also by the screening effect of the $X$-electrons localized in the AlAs layers. This is the difference from work~\cite{Leadley1994}, where a decrease in the role of elastic scattering was achieved only by increasing the width of the spacer. Although it should be noted that in work~\cite{Leadley1994}, with a very large spacer value, the oscillations with $N = 1$ disappeared completely, and in the structures studied here, the magnitude of the peak of this oscillation $N = 1$ was only two times smaller than the peak of the oscillation $N = 2$.

Thus, it has been established that the magnetic field strength at which oscillations of magnetophonon resonance on optical phonons are observed in GaAs structures does not depend on the dimensionality of the system. This fact means that in two-dimensional objects with a GaAs quantum well, resonant electron scattering in the temperature range of 90--180 K occurs on bulk longitudinal optical phonons. Intersubband electron scattering does not affect the qualitative behavior of magnetophonon resonances on optical phonons, and in the temperature range of 90--180 K in a magnetic field of $B \approx 22$ T, inelastic electron scattering prevails over elastic scattering.

\section{MAGNETOPHONON RESONANCE OF RESISTANCE ON ACOUSTIC PHONONS}

\subsection{Introduction}

\begin{figure*}[t]
\centering
\includegraphics[width=1.5\columnwidth]{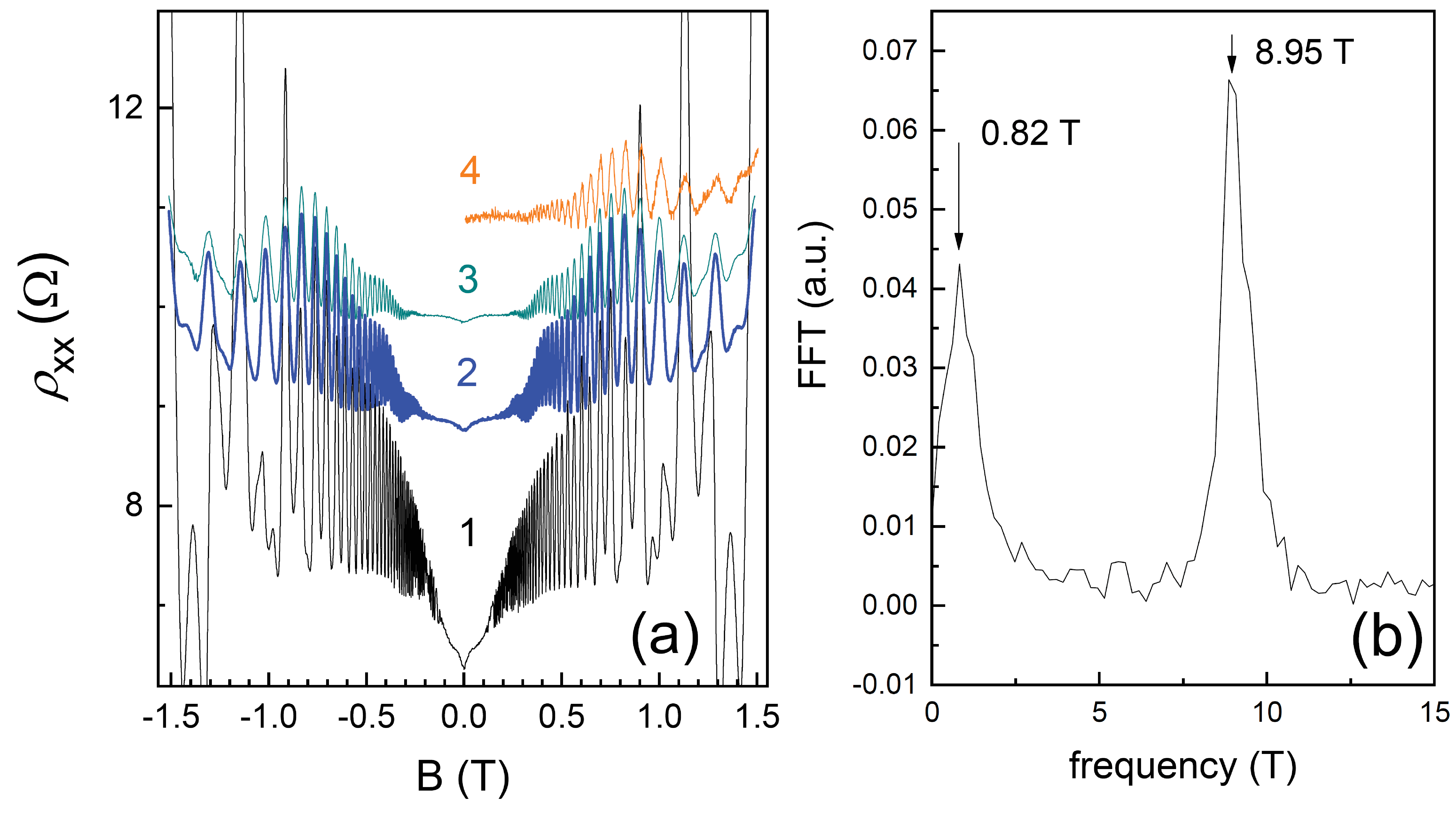} 
\caption{(a) Magnetoresistance $\rho_{xx}$ as a function of magnetic field at different temperatures: 4~K – curve 1, 12~K – curve 2, 16~K – curve 3, 20~K – curve 4
for sample 1. (b) Fourier analysis of the oscillations.}
\label{fig5}
\end{figure*}

In the same samples, but in the temperature range from 10 to 25 K and a stationary magnetic field of up to 2 T, perpendicular to the plane of the sample, oscillations of magnetoresistance are also observed, periodic in the inverse magnetic field. If the position of oscillations in magnetophonon resonance on optical phonons in a magnetic field does not depend on the concentration of charge carriers, then with this type of oscillations their position is determined by the Fermi wave vector $k_{F}$. Absorbing or emitting an acoustic phonon, the electron jumps between Landau levels, and the maximum probability of such transitions is realized during electron backscattering, when the phonon wave vector reaches a value $2k_{F}$ in the 2D plane, i.e., these oscillations can be qualitatively explained by the resonant absorption and emission of acoustic phonon-induced resistance oscillations (PIRO). The condition determining the position of oscillations in a magnetic field for this effect is
\begin{equation}
2k_{F}u=j\omega_{c},
\end{equation}
where $u$ is the velocity of acoustic phonons, $j$ is a positive integer.

Magnetophonon resonance on acoustic phonons was first observed in 2001 in Ref.~\cite{Zudov2001}. It was then studied experimentally and theoretically in Refs.~\cite{Bykov2005,Bykov2009,Hatke2009,Dmitriev2012,Raichev2009,Raichev2010,Bykov2010}. At first, its origin was explained by the resonant interaction of electrons with interface acoustic phonons~\cite{Zudov2001,Bykov2005}. However, calculations~\cite{Raichev2009} then showed that the most probable interaction is the interaction of electrons with bulk acoustic phonons.

The theory of magnetophonon resonance on acoustic phonons in objects with a two-subband spectrum was considered in Ref.~\cite{Raichev2010}, where it was shown that the magnetoresistance of such a system should consist of the sum of the classical magnetoresistance and the quantum contribution associated with the resonant scattering of electrons on phonons within each of the intrasubband, as well as resonant scattering on phonons leading to intersubband transitions.

\subsection{Experimental results and their discussion}

Measurements of the dependence of resistance on a magnetic field of up to 2 T, perpendicular to the plane of the sample, were carried out with a direct current of 1 $\mu$A in the same objects in the temperature range of 4--25 K. Fig.~\ref{fig5} shows the experimental dependences $\rho_{xx}$ on the magnetic field at different temperatures.

As can be seen in Fig.~\ref{fig5}(a), additional slow oscillations are observed against the background of the intersubband oscillations (MISO) pattern, as evidenced by the Fourier analysis presented in Fig.~\ref{fig5}(b). The Shubnikov--de Haas oscillations at these temperatures are shifted to the region of higher magnetic fields. Indeed, the first peak of the Fourier analysis is observed at $B = 8.95$ T, the position of which allows us to calculate the difference carrier concentration in the size quantization subbands $\Delta n = (6.2-1.9) \cdot 10^{11}$ cm$^{-2}$, and knowledge of $\Delta n$ allows us to determine the energy of intersubband separation $\Delta_{SAS}$:
\begin{equation}
\Delta n = \frac{2e}{hc} B = 4.3 \cdot 10^{11}\ \text{cm}^{-2},
\end{equation}
\begin{equation}
\Delta_{SAS} = \frac{\pi \hbar^2}{m^*} \Delta n = 15.5\ \text{meV}.
\end{equation}
The second peak corresponds to the position of the first PIRO oscillation in a magnetic field of 0.82 T. Modulation of the amplitude of intersubband oscillations (MISO) with the frequency (PIRO) demonstrates their significant interference.

\begin{figure}[h]
\centering
\includegraphics[width=\columnwidth]{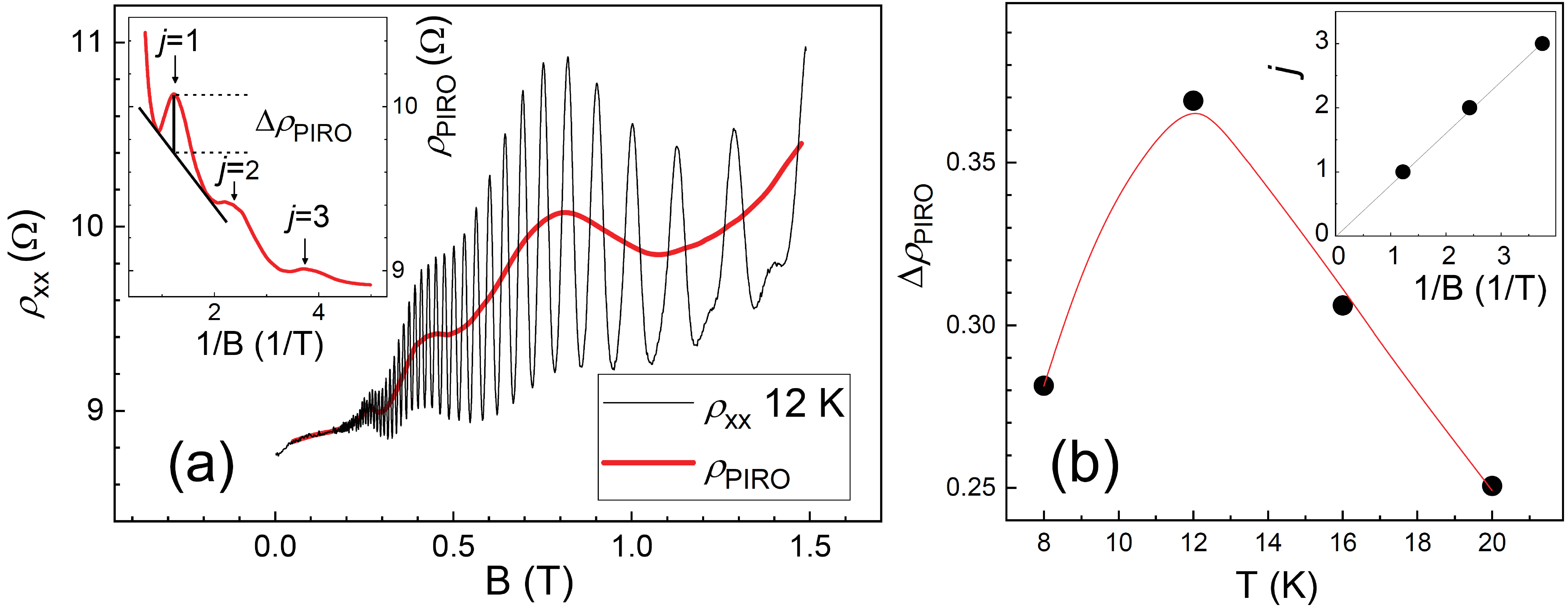} 
\caption{Extraction of slow oscillations for sample 1 at $T$=12~K from the experimental dependence $\rho_{xx}$~($B$). Insert: dependence of the
amplitude of the extracted slow oscillation on 1/$B$, arrows with numbers indicate the oscillation numbers. b – Dependence of the peak
amplitude of the first oscillation on temperature. Inset: dependence of the oscillation number on 1/$B$}
\label{fig6}
\end{figure}

Fig.~\ref{fig6} shows the process of extracting slow oscillations (PIRO) at $T = 12$ K.

The calculations given in Ref.~\cite{Bykov2010} for a similar sample showed that up to 0.8 T the magnetophonon resonance on acoustic phonons (PIRO) is determined by the resonant scattering of electrons of only the first subband (intrasubband) with $n = 6.2 \cdot 10^{11}$ cm$^{-2}$. If this is the case, then the value of the velocity of sound $u$ can be determined from the slope of the linear dependence $j(1/B) = k_F u / \omega_c$ (inset in Fig.~\ref{fig6}(b)). It turned out that the value determined from the slope $u \approx 5.1 \cdot 10^5$ cm/s is close to the value of the longitudinal velocity of sound propagating in the (100) plane in the [110] direction~\cite{Adachi1985}. In addition, as was shown in Ref.~\cite{Hatke2009}, for the amplitude of magnetophonon oscillations on acoustic phonons:
\begin{equation}
\Delta \rho_{PIRO} \propto \tau_{ph}^{-1}(T) \cdot \exp \left( -\frac{2\pi}{\omega_c \tau_q^{ee}(T)} \right),
\end{equation}
where $\tau_{ph}$ is the relaxation time of electrons scattered by acoustic phonons, $\tau_q^{ee}$ is the component of the quantum lifetime due to electron-electron scattering. Since for this sample $1 / \tau_{ph}(T) \propto T^{1.8}$, and $1 / \tau_q^{ee} = 3.8 T^2 / E_F$~\cite{Bykov2010}, in this formula the dependence $\tau_{ph}^{-1}(T)$ determines the temperature increase in the amplitude $\Delta \rho_{PIRO}$, and the factor $\exp[-2\pi / \omega_c \tau_q^{eq}(T)]$ determines its suppression with increasing temperature, which is observed in Fig.~\ref{fig6}(b) for the first oscillation.

We believe that the position of the maxima of slow oscillations in a magnetic field, the magnitude of the acoustic wave velocity determined in this experiment, and the temperature dependence of the maximum amplitude of the first peak confirm that the observed slow oscillations are associated with resonant scattering of electrons on bulk acoustic phonons.

The presented experimental results are close to the results of work~\cite{Bykov2010}, carried out on a similar sample. On sample 2 with a wide well, we were unable to distinguish between intersubband oscillations and PIRO.

Thus, the magnetophonon resonance on acoustic phonons observed in the temperature range of 8--20 K in a sample with a single quantum well and two subbands of size quantization is determined by the resonant scattering of electrons on longitudinal acoustic phonons in the first subband. The occurrence of these oscillations causes the interference contribution of MISO-PIRO to the magnetoresistance.

\section*{ACKNOWLEDGMENTS}

The authors are grateful to Yu. M. Galperin and R. V. Parfeniev for discussing the results and carefully reading the manuscript.

\end{document}